\documentclass[journal, twoside, 10pt]{IEEEtran}
\usepackage{graphicx}
\usepackage{amssymb}
\usepackage{amsmath}
\usepackage{calc}
\usepackage{array}
\usepackage{subfigure}
\usepackage{graphicx}
\usepackage{algorithm}
\usepackage{amsthm}
\usepackage{algpseudocode}

\theoremstyle{definition}

\newcolumntype{M}[1]{>{\centering\arraybackslash}m{#1}}
\newcolumntype{N}{@{}m{0pt}@{}}

\newcommand\blfootnote[1]{
  \begingroup
  \renewcommand\thefootnote{}\footnote{#1}
  \addtocounter{footnote}{-1}
  \endgroup
}

\begin{document}
\title{Information-Theoretic Privacy in Federated Submodel learning} 

\author{
  \IEEEauthorblockN{Minchul Kim,~\IEEEmembership{Student Member,~IEEE} and Jungwoo Lee,~\IEEEmembership{Senior Member,~IEEE}}
\vspace{-5mm}
}

\maketitle

\begin{abstract}

We consider information-theoretic privacy in \textit{federated submodel learning}, where a global server has multiple submodels.
Compared to the privacy considered in the conventional federated submodel learning where secure aggregation is adopted for ensuring privacy, information-theoretic privacy provides the stronger protection on submodel selection by the local machine.
We propose an achievable scheme that partially adopts the conventional \textit{private information retrieval} (PIR) scheme that achieves the minimum amount of download.
With respect to computation and communication overhead, we compare the achievable scheme with a naïve approach for federated submodel learning with information-theoretic privacy.
\end{abstract}


\section{Introduction}
\label{intro}
\blfootnote{


M. Kim and J. Lee are with the Communications and Machine Learning Lab., Department of Electrical and Computer Engineering, Seoul National University, Seoul, 08826, Korea (e-mail: kmc1222@cml.snu.ac.kr junglee@snu.ac.kr).
}

As machine learning scales larger, the learning task cannot be handled by a single machine.
As a result, learning with a number of distributed local machines has been studied for speeding up the learning process.
On the other hand, as personal training data can be transferred to a global server, local data protection (security) against the global server has been considered as a significant issue.
For this data security, \textit{federated learning} \cite{FL} has been proposed, where local machines only upload their local updates to the global server.

In \cite{FSL}, federated submodel learning has been proposed, where a global server stores a number of submodels and each local machine selectively trains a subset of submodels.
Due to this submodel selection, privacy on selecting submodels should also be considered in federated submodel learning, as well as the data security.
In \cite{FSL}, \textit{secure aggregation} \cite{SA} has been adopted for the privacy on submodel selection.

In this letter, we consider information-theoretic privacy in federated submodel learning.
In \cite{FSL}, a global server may take advantage of the fact that the submodels that are not aggregated by secure aggregation are not chosen by local machines.
On the other hand, information-theoretic privacy implies that a global machine cannot differentiate a chosen submodel from all the other submodels.
Therefore, the privacy considered in our work is stronger than that of \cite{FSL}.

For information-theoretic privacy in federated submodel learning, we consider multiple non-colluding global servers which do not communicate with each other.
Compared to the single global server case in \cite{FSL}, the assumption of multiple global servers may be a restriction.
However, unlike \cite{FSL}, there is no need to aggregate local machines for ensuring the privacy on submodel selection.
That is, a local machine does not need to wait for the other local machines for ensuring the privacy.
Considering multiple non-colluding global servers in practical scenarios, one local machine may participate in several independent groups (e.g., different companies) for federated learning and there may be a common set of submodels across the groups.
Since these groups do not communicate with each other, local machine aggregation across the groups is usually unavailable.
Our work can be applied to this kind of scenarios and ensures privacy on the submodel selection by one local machine.

We propose an achievable scheme that exploits the conventional \textit{private information retrieval} (PIR) \cite{PIR} technique.
In PIR literature, information-theoretic privacy has been considered only for downloading.
On the other hand, our achievable scheme also handles the uploading process of federated learning.
Specifically, in the download phase of our achievable scheme, we adopt a PIR scheme that achieves the minimum amount of download for non-colluding servers \cite{HSun}.
With respect to communication and computation overhead, we compare our achievable scheme with an naive approach for information-theoretic privacy in federated submodel learning.
Compared to \cite{FSL}, since we assume a single local machine without aggregation, comparison with secure aggregation scheme proposed in \cite{FSL} is unfeasible.
We prove that our achievable scheme ensures privacy and characterize the lower bound of overheads.

\textit{Notation }: An integer set from $1$ to $N$ is denoted by $[N]$ and a set from $N$ to $M$ is denoted by $[N:M]$.

\section{System Model}
\label{systemmod}
We assume that there are $r$ submodels and each submodel identically has $s$ parameters. 
The parameters of all submodels are aggregated in a matrix $\text{B}_t\in\mathbb{F}_q^{r \times s}$ whose $r$ rows $\{\text{B}_{t,1}, \text{B}_{t,2}, \cdots \text{B}_{t,r}\}$ denote the parameters of $r$ submodels.
The index $t$ in $\text{B}_t$ denotes the current iteration.
That is, there were $t-1$ model updates in the parameter matrix $\text{B}_t$ before.
For the privacy of local machines, the parameter matrix $\text{B}_t$ is encoded into $\tilde{\text{B}}_t$ whose $r$ rows are denoted by $\{\tilde{\text{B}}_{t,1},\tilde{\text{B}}_{t,2}, \cdots, \tilde{\text{B}}_{t,r}\}$.
The detail of encoding will be explained in the next section.

As global servers where the local updates are aggregated, we assume that there are $N$ non-colluding databases $\{\text{DB}_{i}\}_{i=1}^{N}$ who do not communicate with each other.
All of the databases share the matrix $\tilde{\text{B}}_t$ in a replication-based way.
Whereas the databases share the parameter matrix $\tilde{\text{B}}_t$, there is an exclusive data that is stored only in each database, which will be specified in the next section.
We sequentially denote the exclusive data for each database by $c_{t-1,1}, c_{t-1,2}, \cdots, c_{t-1,N}$.
Since the databases do not collude each other, each $DB_i$ cannot know the other exclusive data $\{c_{t-1,j}\}_{j\in[N]\setminus i}$.

We denote the local machine at iteration $t$ by $L_t$.
Note that each iteration is occupied by one local machine.
Among $r$ submodels, we assume that $L_t$ wants to update a specific submodel whose parameters are given by $\text{B}_{t,d}$, the $d$th row of $\text{B}_t$.
At the iteration $t$, there are three operations of $L_t$: downloading, updating, and uploading, which will be specified in the next section.
The privacy of $L_t$ implies that the index $d$ is concealed from the databases after the aforementioned three operations. 
For each database $DB_i$, this privacy constraint can be expressed as below.
\begin{gather}
\label{constraint}
      I(d;Q_{t,i},\mathcal{U}_{t,i},S_{t,i})=0, 
\end{gather}
where $Q_{t,i}$, $\mathcal{U}_{t,i}$, and $S_{t,i}$ denote the queries that $L_t$ sends to $DB_i$, local data uploaded from $L_t$ to $DB_i$, and $\text{B}_{t} \cup c_{t-1,i}$, respectively.

\begin{figure}[t]
    \centerline{\includegraphics[width=7cm]{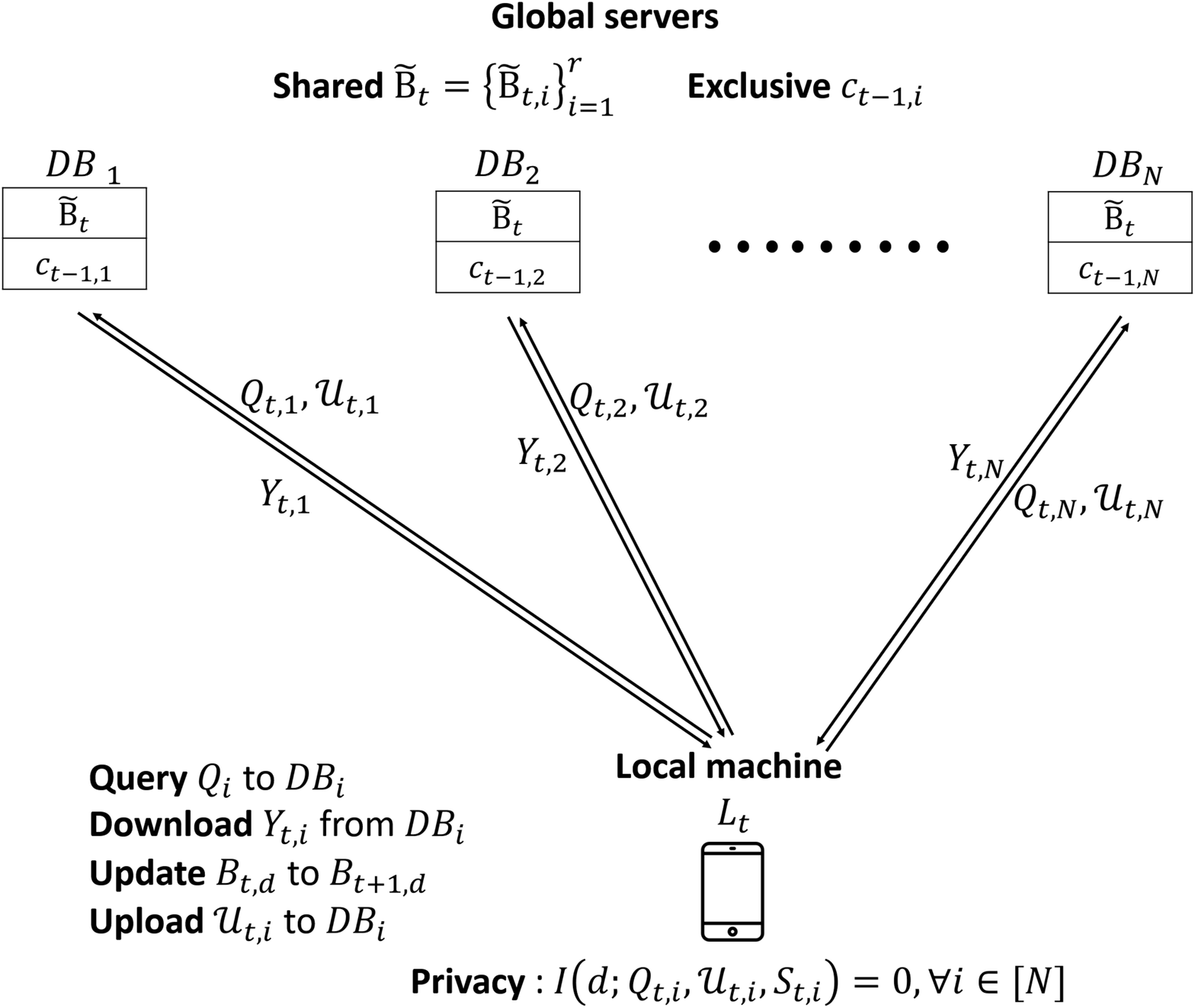}}
    \caption{System model.}
    \label{sysmod}
    \vspace{0mm}
\end{figure}

With respect to overhead, we consider the communication overhead and computation overhead for ensuring the privacy at the iteration $t$.
The communication overhead includes the amount of download and upload.
On the other hand, the computation overhead includes the amount of computation for encoding, decoding, training, and updating at the local machine $L_t$.
We depict the system model in Fig. \ref{sysmod}.

\section{Achievable scheme}
\label{achi}
At first, we explain the detail of encoded parameter matrix $\tilde{\text{B}}_t$ and the exclusive data $c_{t-1,i}$ for each database $DB_i$.
Suppose that the local machine $L_{t-1}$ at the previous iteration $t-1$ desired the parameter vector $\text{B}_{t-1,p}$.
As a result, after the iteration $t-1$, the parameter vector $\text{B}_{t-1,p}$ would be updated into $\text{B}_{t,p}$.
We denote the difference between the vectors $\text{B}_{t,p}$ and $\text{B}_{t-1,p}$ by $\Delta_{t-1}$.
That is, we have
\begin{gather}
\label{t-1}
    \text{B}_{t,p}=\text{B}_{t-1,p}+\Delta_{t-1}.
\end{gather}
Note that $\text{B}_{t,l}=\text{B}_{t-1,l}, \forall l\in[r]\setminus p$.
The difference vector $\Delta_{t-1}$ also applies to the other parameter vectors, with coefficient $\alpha_{t-1,l}$ for each $l\in[r]\setminus p$, where $\alpha_{t-1,p}=1$.
That is, a scaled vector $\alpha_{t-1,l}\Delta_{t-1}$ is added when encoding $\tilde{\text{B}}_{t,l}$.
Note that the coefficients $\{\alpha_{t-1,1}, \alpha_{t-1,2}, \cdots, \alpha_{t-1,r}\}$ are distinct from each other.
As a result, the vector $\tilde{\text{B}}_{t,l}$ is given as
\begin{gather}
\label{param}
      \tilde{\text{B}}_{t,l}=\text{B}_{t-1,l}+\alpha_{t-1,l}\Delta_{t-1}, \forall l\in[r].
\end{gather}

We now explain the exclusive data $c_{t-1,i}$ for each $i\in[N]$.
All of the exclusive data $c_{t-1,1}, c_{t-1,2}, \cdots, c_{t-1,N}$ across the databases can be decoded into a message $M_{t-1}$.
The message $M_{t-1}$ contains the two information elements: the difference vector $\Delta_{t-1}$ and a coefficient vector $\alpha_{t-1}=\{\alpha_{t-1,1}, \alpha_{t-1,2}, \cdots, \alpha_{t-1,r}\}$, thus implying that
\begin{gather}
\label{msg}
    M_{t-1}=\{\alpha_{t-1}, \Delta_{t-1}\}.
\end{gather}
Note that the $M_{t-1}$ has $r+s$ bits.
Accordingly, the encoding of $M_{t-1}$ is done with a non-systematic $(r+s,r+s)$-MDS code, by the local machine $L_{t-1}$, at the previous iteration $t-1$.


After encoding, $r+s$ encoded bits are equally partitioned into the exclusive data $c_{t-1,1}, c_{t-1,2}, \cdots, c_{t-1,N}$, each of which $(r+s)/N$ bits long.
Since the data $c_{t-1,i}$ is exclusive to $DB_i$ and the databases do not collude each other, each database $DB_i$ cannot infer $M_{t-1}$ in spite of having $c_{t-1,i}$.

As explained in Section \ref{systemmod}, an achievable scheme for the aforementioned system model considers the download, update, and upload, with respect to the local machine $L_t$ at the iteration $t$.
We sequentially explain the download phase, update phase, and upload phase.

\subsection{Download phase}
\label{dphase}
There are two steps in download phase: exclusive data download and shared data download.
The first step is to download the exclusive data $c_{t-1,i}$ from each database $DB_i$.
Since each exclusive data is a partition of MDS-coded bits,  downloading it from each database does not offer any information on the index $d$ to each database in terms of privacy.
Accordingly, in the first step, the local machine $L_t$ only downloads $c_{t-1,i}$ from $DB_i$ without any further download to ensure privacy.
As a result, the amount of download in the first step is $r+s$ bits.

The second step of download phase is to download the encoded parameter vector $\tilde{\text{B}}_{t,d}$ from the databases while ensuring privacy.
As explained in the Section \ref{systemmod}, the vector $\tilde{\text{B}}_{t,d}$ is replicated in every database, unlike the exclusive data $c_{t,i}$ is stored only in $DB_i$.
Furthermore, the local machine $L_t$ aim to conceal the index $d$ from the databases.
As a result, in the second step, $L_t$ downloads $\tilde{\text{B}}_{t,d}$ with the conventional PIR scheme \cite{HSun} which minimizes the amount of download for replication-based databases, while ensuring the privacy on $d$.
From \cite{HSun}, the minimal amount of downloaded bits is given by $(1+\frac{1}{N}+\frac{1}{N^2}+\cdots+\frac{1}{N^{r-1}})s=(1+\beta)s$.
Note that $s$ bits correspond to the desired submodel $\tilde{\text{B}}_{t,d}$ and $\beta s$ bits correspond to the portion of the undeisred submodels $\{\tilde{\text{B}}_{t,l}| l\in[r]\setminus d\}$, which is additionally downloaded for ensuring the privacy.

\subsection{Update phase}
\label{udphase}
In the update phase, the local machine $L_t$ updates the parameter vector $\text{B}_{t,d}$ into $\text{B}_{t+1,d}$ for the next iteration $t+1$.
At first, for obtaining $\text{B}_{t,d}$ from $\tilde{\text{B}}_{t,d}$, $L_t$ decodes the downloaded exclusive data $c_{t-1,1}, c_{t-1,2}, \cdots, c_{t-1,N}$ into the message $M_{t-1}=\{\alpha_{t-1},\Delta_{t-1}\}$.
Subsequently, $L_t$ obtains $\text{B}_{t,d}$ from $\tilde{\text{B}}_{t,d}=\text{B}_{t-1,d}+\alpha_{t-1,l}\Delta_{t-1}$.
After that, $L_t$ trains the desired submodel of the parameters $\text{B}_{t,d}$.
Note that $L_t$ does not train any undesired submodel.

After the training, $L_t$ obtains a new difference vector $\Delta_{t}=\text{B}_{t+1,d}-\text{B}_{t,d}$ and generates a new coefficient vector $\alpha_t=\{\alpha_{t,1},\alpha_{t,2}, \cdots, \alpha_{t,r}\}$.
As the coefficient $\alpha_{t-1,p}$ was $1$ in the iteration $t-1$, the coefficient $\alpha_{t,d}$ is $1$ in the iteration $t$.
The other coefficients of $\alpha_t$ are randomly chosen and distinct from each other.
Subsequently, $L_t$ encodes the message $M_{t}=\{\alpha_t,\Delta_t\}$ with a $(r+s,r+s)$-MDS code, and partitions the encoded bits into exclusive data $c_{t,1}, c_{t,2}, \cdots, c_{t,N}$.
After encoding, $L_t$ computes $r$ linear combinations of $\Delta_{t-1}$, $\Delta_t$, $\alpha_{t-1}$, and  $\alpha_t$, which are to be uploaded for each submodel.
We specify the linear combinations in the upload phase.

We now characterize the computation overhead.
The computation overhead for training the desired submodel is denoted by $f(s)$ since the model structure is unspecified.
The overheads for encoding and decoding of a $(r+s,r+s)$-MDS code are $\mathbb{O}((r+s)^2)$ and $\mathbb{O}((r+s)^3)$, respectively.
If a Vandermonde matrix is used for encoding, the decoding overhead can be reduced to $\mathbb{O}((r+s)^2)$.

\subsection{Upload phase}
\label{uphase}
Similar to the download phase, there are two steps in the upload phase: exclusive data upload and shared data upload.
In the first step, the local machine $L_t$ uploads the exclusive data $c_{t,i}$ to each $DB_i$ where the overhead is $r+s$ bits. 
As explained in the download phase, since each exclusive data is a partition of MDS-coded bits, uploading it to each database does not offer any information on the index $d$ to each database in terms of privacy.

In the second step, for each database $DB_i$, the local machine $L_t$ uploads the aforementioned $r$ linear combinations of $\Delta_{t-1}$, $\Delta_t$, $\alpha_{t-1}$, and $\alpha_t$, which are denoted by $U_t=\{U_{t,1},U_{t,2},\cdots,U_{t,r}\}$ and specified as follows.
\begin{gather}
\label{Ui}
    U_{t,l}=\alpha_{t,l} \Delta_t - \alpha_{t-1,l} \Delta_{t-1} , \forall l\in[r]\setminus p.\\
\label{Ui2}
    U_{t,p}=\alpha_{t,p} \Delta_t, l=p.
\end{gather}
Recall that $\alpha_{t-1,p}=\alpha_{t,d}=1$.
After uploading, each $DB_i$ adds $U_{t,l}$ to $\tilde{\text{B}}_{t,l}$.
From (\ref{param}), (\ref{Ui}), and (\ref{Ui2}), this equals to the encoded vector for the iteration $t+1$, which is given by  
\begin{gather*}
    \tilde{\text{B}}_{t+1,l}=\text{B}_{t,l}+\alpha_{t,l}\Delta_{t}, \forall l\in[r].
\end{gather*}
Note that $\text{B}_{t,l}=\text{B}_{t-1,l}, \forall l\in[r]\setminus p$.
Since $L_t$ uploads $r$ linear combinations for each database, the overhead of the second step is $rsN$ bits, thus implying that the communication overhead in the upload phase is given by $rsN+r+s$ bits.
In Algorithm \ref{alg}, the overall process of our achievable scheme is summarized with respect to $L_t$.


\begin{algorithm}[t]
\caption{The overall process of $L_{t}$} \label{alg}
\begin{algorithmic}
\Procedure{download phase}{}
\For {$i=1 \to N$}
\State Step 1. Download $c_{t-1,i}$ from $DB_i$
\State Step 2. PIR scheme \cite{HSun} for obtaining $\tilde{\text{B}}_{t,d}$
\EndFor
\EndProcedure
\Procedure{update phase}{}
\State Decode : $\{c_{t-1,i}|i\in[N]\} \to M_{t-1}$
\State Decode : $\{M_{t-1},\tilde{\text{B}}_{t,d}\} \to \text{B}_{t,d}$
\State Learn : $\text{B}_{t,d} \to \text{B}_{t+1,d}$
\State Update : $M_{t-1} \to M_t$
\State Compute : $\{\text{B}_{t,d}, M_{t-1}, M_t\} \to U_t$
\State Encode : $M_t \to \{c_{t,i}|i\in[N]\}$
\EndProcedure
\Procedure{upload phase}{}
\For {$i=1 \to N$}
\State Step 1. Upload $c_{t,i}$ to $DB_i$
\State Step 2. Upload $U_t$ to $DB_i$
\EndFor
\EndProcedure
\end{algorithmic}
\end{algorithm}

\section{Overhead comparsion}
\label{OH}
In this section, we compare the overhead of the achievable scheme explained in the previous section with respect to the naive approach.

We first explain the naive approach.
In the download phase, the local machine $L_t$ downloads whole $r$ submodels to ensure the privacy, thus implying that the overhead is $rs$ bits.
We assume that $L_t$ equally downloads $rs/N$ bits from each database.
In the update phase, $L_t$ trains whole $r$ submodels.
Note that this is due to the two constraints: one is for the privacy and the other is for the continuity of the federated learning process.
For the privacy, the amount of update for each submodel need to be identical at each database.
On the other hand, for the continuity, all the updates of the submodels should be correct.
Therefore, we assume that the naive approach for satisfying both of the constraints is to download and train all $r$ submodels.
In the upload phase, for each database, $L_t$ uploads the update for whole $r$ submodels, thus implying that the communication overhead is $rsN$ bits.

We now compare the naive approach with our achievable scheme.
We first compare the communication overhead.
For the download, our achievable scheme downloads $r+(2+\beta)s$ bits, whereas the naive approach downloads $rs$ bits.
Therefore, in many practical scenarios where $r$ and $s$ are large, our achievable scheme outperforms the naive approach with respect to the amount of download.
For the upload, our achievable scheme uploads $rsN+r+s$ bits, whereas the naive approach uploads $rsN$ bits.
That is, the naive approach outperforms our achievable scheme with respect to the amount of upload.
The overall communication overheads are $2r+(3+\beta+rN)s$ and $rs(N+1)$ bits, respectively.
Since the difference between the two overheads is given by $(2r+(3-r+\beta)s)$ bits, our achievable scheme generally outperforms the naive approach for $r>3$ with respect to the overall communication overhead.

We now compare the computation overhead.
As explained in the previous section, there are two kinds of computational overhead in our achievable scheme. 
One is the training overhead and the other is encoding/decoding overhead.
The training overhead $f(s)$ implies the computation for training one desired submodel.
The encoding/decoding overhead equals to two times the $\mathbb{O}((r+s)^2)$ operations.
On the other hand, in the naive approach, only training overhead is considered where the local machine $L_t$ need to train whole $r$ submodels.
Note that each submodels has $s$ parameters and therefore the computation overhead is given by $rf(s)$.
Therefore, in practical, in spite of encoding/decoding overhead, it is obvious that our achievable scheme outperforms the naive approach with respect to the computation overhead.
We summarize the comparison in Table 1.

\begin{table} [t]
\centering
\caption{Overhead comparison}
\label{table}
\begin{tabular}{ |M{1.5cm}|M{2.5cm}|M{2cm}|N } 
 \hline
 &proposed & naive &\\[4pt]
 \hline
 computation&$f(s)$+$2\times\mathbb{O}((r+s)^2)$ & $rf(s)$ &\\[4pt]
 \hline
 download&$r+(2+\beta)s$ & $rs$ &\\[4pt]
 \hline
 upload&$rsN+r+s$  & $rsN$ &\\[4pt]
 \hline
 overall&$2r+(3+\beta+rN)s$ & $rs(N+1)$ &\\[4pt]
 \hline
\end{tabular}
\end{table}

\section{Privacy proof and optimality analysis}
\label{proof}

\subsection{Privacy proof}
\label{pp}
For the privacy proof, we show that the constraint (\ref{constraint}) is satisfied for every database.
By the chain rule, the constraint (\ref{constraint}) for $DB_i$ becomes as follows.
\begin{gather*}
    I(d;Q_{t,i})+I(d;\text{S}_{t,i}|Q_{t,i})+I(d;\mathcal{U}_{t,i}|Q_{t,i},\text{S}_{t,i})=0.
\end{gather*}
We show that each of three terms equals to $0$.
We first show that $\text{S}_{t,i}=\text{B}_t \cup c_{t-1,i}$ is independent to $d$.
Since the local machine $L_t$ determines $d$ before downloading any parameter, $L_t$ does not have any information about $\text{S}_{t,i}$ when determining $d$, thus implying that $I(d;\text{S}_{t,i}|Q_{t,i})=0$.

We now show that the third term equals to $0$. 
The uploaded data $\mathcal{U}_{t,i}$ has two elements: $c_{t,i}$ and $U_t$.
Recalling that the exclusive data $c_{t,i}$ for $DB_i$ is a partition of non-systematic MDS-coded $M_{t}=\{\alpha_t,\Delta_t\}$, $DB_i$ cannot specify $M_t$ by having $c_{t,i}$.
For $L_t$, we can vectorize (\ref{Ui}) as follows.
\begin{gather}
\label{LCs}
    U_t=\begin{bmatrix}
\alpha_{t,1} & \alpha_{t-1,1} \\ \alpha_{t,2} & \alpha_{t-1,2} \\ \vdots & \vdots \\ \alpha_{t,r} & \alpha_{t-1,r}
\end{bmatrix}\begin{bmatrix} \Delta_{t} \\ -\Delta_{t-1} \end{bmatrix}.
\end{gather}
By mathematical induction, we show that (\ref{LCs}) is underdetermined.
For the first iteration $t=1$, (\ref{LCs}) would be $U_1 = \alpha_1^T\Delta_1$.
Since both of $\alpha_1$ and $\Delta_1$ are generated by the first local machine $L_1$, they are unknown to the databases, thus implying that every database cannot solve the system $U_1 = \alpha_1^T\Delta_1$.
For the iteration $t-1$, we assume that the system $U_{t-1}=\alpha_{t-1}^T\Delta_{t-1}-\alpha_{t-2}^T\Delta_{t-2}$ is underdetermined, thus implying that $\alpha_{t-1}$ and $\Delta_{t-1}$ remain unknown to the databases.
As a result, at iteration $t$, the vectors $\alpha_{t-1}$ and $\Delta_{t-1}$ are unknown from the previous iteration and the vectors $\alpha_{t}$ and $\Delta_{t}$ are generated by the local machine $L_t$.
Since all of the vectors $\alpha_t$, $\alpha_{t-1}$, $\Delta_t$, and $\Delta_{t-1}$ are unknown to each database $DB_i$, the system (\ref{LCs}) is also underdetermined.
Therefore, the uploaded data $\mathcal{U}_{t,i}$ does not give any information on $d$ to $DB_i$, thus implying that $I(d;\mathcal{U}_{t,i}|Q_{t,i},\text{S}_{t,i})=0$.

We now show that the queries $Q_{t,i}$ does not give any information about $d$.
In our achievable scheme, there are four kinds of queries : downloading $c_{t-1,i}$ and $\tilde{\text{B}}_{t,d}$, uploading $c_{t,i}$ and $U_t$.
According to the queries for downloading $c_{t-1,i}$, and uploading $c_{t,i}$ and $U_t$, $L_t$ merely downloads and uploads whole content of $c_{t-1,i}$, $c_{t,i}$, and $U_t$ which is independent to $d$.
Therefore, it is obvious that these three kinds of queries does not give any information on $d$.

We now consider the queries for downloading $\tilde{\text{B}}_{t,d}$.
Recall that we adopt the conventional PIR scheme \cite{HSun} for downloading parameter vector $\tilde{\text{B}}_{t,d}$, which has been proven to ensure privacy, thus implying that $I(d;Q_{t,i})=0$.$\square$

\subsection{Optimality analysis}
In this section, we characterize the lower bound of overheads.
As for download overhead, it has been proven that the PIR scheme in \cite{HSun} achieves the minimum amount of download, $(1+\beta)s$.
Compared to the lower bound, our achievable scheme requires $r+s$ more bits for download.
As for computation overhead, it is obvious that the minimum is $f(s)$, which is the same as for the case where the privacy does not have to be considered.
Compared to the lower bound, our achievable scheme requires $2\mathbb{O}((r+s)^2)$ more bits for encoding and decoding.

We now show that the lower bound for upload overhead equals $rsN$ bits.
That is, for each database, the amount of uploaded bits should be more than $rs$ bits.
For the desired submodel, the local machine uploads $s$ bits for parameter update.
Since there should be no difference among the submodels, the local machine should upload $s$ bits for the other undesired submodels.
Therefore, the minimum amount of uploaded bits for each database equals to $rs$ bits, as we claim.
Compared to the lower bound, our achievable scheme requires $r+s$ more bits for uploading.


\begin{thebibliography}{00}

\bibitem{FL}
 B. McMahan and D. Ramage. (Apr. 2017). ``Federated Learning: Collaborative Machine Learning Without Centralized Training
Data," [Online]. Available: https://ai.googleblog.com/2017/04/federatedlearning-collaborative.html

\bibitem{FSL}
 C. Niu, F. Wu, S. Tang, L. Hua, R. Jia, C. Lv, Z. Wu, and G. Chen. ``Secure federated submodel learning," \emph{arXiv preprint arXiv:1911.02254}, 2019.

\bibitem{SA}
   K. Bonawitz, V. Ivanov, B. Kreuter, A. Marcedone, H. B. McMahan, S. Patel, D. Ramage, A. Segal, and K. Seth, ``Practical secure aggregation for privacy-preserving machine learning," \emph{Proc. of CCS}, 2017, pp. 1175–1191.

\bibitem{PIR}
 B. Chor, E. Kushilevitz, O. Goldreich, and M. Sudan, ``Private information
retrieval,'' in
\emph{Journal of the ACM}, 45(6):965-981, 1998.

\bibitem{HSun}
 H. Sun and S. A. Jafar, ``The Capacity of Private Information Retrieval,'' in
\emph{IEEE Transactions on Information Theory}, vol. 63, no. 7, pp. 4075-4088, Jul. 2017.

\end{thebibliography}
\end{document}